\documentclass[%
 reprint,
superscriptaddress,
 bibnotes,
 amsmath,amssymb,
 aps,
 prb,
 citeautoscript,
]{revtex4-2}
\usepackage{graphicx}% Include figure files
\usepackage{dcolumn}% Align table columns on decimal point
\usepackage{bm}% bold math
\usepackage{xcolor} %colored text
\usepackage{hyperref}% add hypertext capabilities
\usepackage{placeins}

\begin{document}
\preprint{APS/123-QED}

\title{Observation of vanishing charge dispersion of a nearly-open superconducting island}

\author{Arno Bargerbos}
\email[]{a.bargerbos@tudelft.nl}
\affiliation{QuTech and Kavli Institute of Nanoscience, Delft University of Technology, 2600 GA Delft, The Netherlands}
\author{Willemijn Uilhoorn}
\affiliation{QuTech and Kavli Institute of Nanoscience, Delft University of Technology, 2600 GA Delft, The Netherlands}
\author{Chung-Kai Yang}
\affiliation{Microsoft Quantum Lab Delft, 2600 GA Delft, The Netherlands}
\author{Peter Krogstrup}
\affiliation{Microsoft Quantum Materials Lab and Center for Quantum Devices, Niels Bohr Institute, University of Copenhagen, Kanalvej 7, 2800 Kongens Lyngby, Denmark}
\author{Leo P. Kouwenhoven}
\affiliation{QuTech and Kavli Institute of Nanoscience, Delft University of Technology, 2600 GA Delft, The Netherlands}
\affiliation{Microsoft Quantum Lab Delft, 2600 GA Delft, The Netherlands}
\author{Gijs de Lange}
\affiliation{Microsoft Quantum Lab Delft, 2600 GA Delft, The Netherlands}
\author{Bernard van Heck}
\affiliation{Microsoft Quantum Lab Delft, 2600 GA Delft, The Netherlands}
\author{Angela Kou}
\affiliation{Microsoft Quantum Lab Delft, 2600 GA Delft, The Netherlands}

\date{\today}% It is always \today, today,
       % but any date may be explicitly specified

\begin{abstract}
Isolation from the environment determines the extent to which charge is confined on an island, which manifests as Coulomb oscillations such as charge dispersion. We investigate the charge dispersion of a nanowire transmon hosting a quantum dot in the junction. We observe rapid suppression of the charge dispersion with increasing junction transparency, consistent with the predicted scaling law which incorporates two branches of the Josephson potential. We find improved qubit coherence times at the point of highest suppression, suggesting novel approaches for building charge-insensitive qubits.
\end{abstract}

\maketitle
The manipulation of single charge carriers has been one of the most important advances in condensed matter physics, enabling a wide range of nanoelectronic technology in areas such as detection, thermometry, and metrology \cite{Likharev1999, Komiyama2000, Lu2003, Meschke2011, Pekola2013}. The control of single charge carriers is made possible by the quantization of charge on mesoscopic islands well-isolated from the environment. Charge quantization manifests as Coulomb oscillations: periodic dependence of the system's observables reflecting the energy cost of adding an additional charge to the system. As the coupling strength to the environment increases, quantum fluctuations progressively delocalize the charge, suppressing Coulomb oscillations. In normal state conductors, it is well-known that this suppression occurs through single-electron tunneling \cite{Flensberg1993, Matveev1995, Aleiner1998, Nazarov1999, Kouwenhoven1991, Molenkamp1995, Joyez1997, Zaikin1999, Jezouin2016}. 

In the case of superconducting islands, the coupling to the environment instead occurs via coherent Cooper pair tunneling. In conventional tunnel junctions, the latter is mediated by a large number of weakly transmitting transport channels, characterized by the Josephson energy $E_{\mathrm{J}}$. In this case, the size of the charge dispersion depends only on the ratio between the charging energy $E_{\mathrm{c}}$ and $E_{\mathrm{J}}$, as illustrated by the Cooper pair box ($E_{\mathrm{J}}/E_{\mathrm{c}} \approx 1$) \cite{Bouchiat1998} and the transmon ($E_{\mathrm{J}}/E_{\mathrm{c}} \gg 1$) \cite{Koch2007}. The Cooper pair box has large charge dispersion, whereas for the transmon charge dispersion is exponentially suppressed in the ratio $E_{\mathrm{J}}/E_{\mathrm{c}}$ \cite{Koch2007, Averin1985}. This behaviour originates from quantum tunneling of the superconducting phase difference $\phi$ below the Josephson potential barrier connecting two energy minima at $\phi=0,2\pi$.

The situation becomes more interesting if the Cooper pair tunneling is mediated by a single transport channel with high transparency \cite{Averin1999}. In this limit, the energy spectrum of the Josephson junction is characterized by a narrowly avoided level crossing at $\phi=\pi$, and imaginary-time Landau-Zener (ITLZ) tunneling \cite{Averin1999, Pikulin2019} acts to prevent quantum tunneling trajectories from reaching the energy minimum at $2\pi$. The charge dispersion of the superconducting island then vanishes completely as the transparency approaches unity. While some weak suppression of Coulomb oscillations has been observed in weak-links \cite{Lorenz2018, Proutski2019}, the effect of ITLZ tunneling on charging effects has eluded experimental verification because of the stringent requirements for ballistic Josephson junctions. However, recent advances in nanofabrication and nanowire growth \cite{Krogstrup2015} have enabled the development of superconductor-semiconductor-superconductor junctions with a small number of highly transmitting modes \cite{VanWoerkom2017,Goffman2017}. Experiments in such devices have detected a single mode with nearly perfect transmission, attributed to resonant tunneling through an accidental quantum dot in the junction \cite{Spanton2017, Hart2019a}. Charge-sensitive devices connected to reservoirs via quantum-dot-based junctions are thus ideal for the investigation of near-ballistic Josephson junction behaviour.

In this Letter, we experimentally investigate the charge dispersion of a superconducting island connected to a reservoir via a semiconducting weak-link hosting a quantum dot. The device constitutes an offset-charge-sensitive (OCS) nanowire transmon, also known as a gatemon \cite{DeLange2015,Larsen2015, Kringhoj2018, Serniak2019}. By in-situ tuning of the transparency of the weak-link using an electrostatic gate, we observe its charge dispersion decrease by almost two decades in frequency at a rate far exceeding exponential suppression in $E_{\mathrm{J}}/E_{\mathrm{c}}$. The observed gate dependence of the charge dispersion is modeled by tunneling through a resonant level, incorporating the effect of ITLZ tunneling. This model agrees well with the measured suppression of charge dispersion, suggesting near-unity junction transparency. Finally, we observe improved qubit coherence times $T_2^*$ and $T_2^{\mathrm{echo}}$ in regions of vanishing charge dispersion, which reflects the strong reduction in the charge sensitivity of the qubit. 

\begin{figure}[t!]
\includegraphics[width=8.6cm]{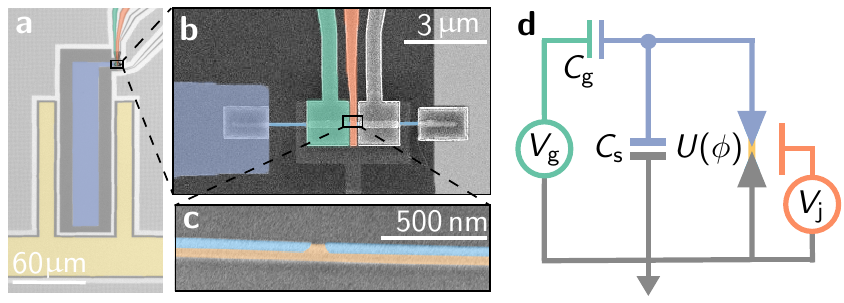}
\caption{\label{fig:fig1} (a) False-color optical microscope image of the qubit. It consists of an island (purple) capacitively coupled to the ground plane (grey) and a CPW resonator (yellow). (b) Scanning electron micrograph (SEM) of the InAs-Al nanowire connecting the island (left) to the ground plane (right). Its weak-link is tuned by the junction gate (red), while the island gate (green) tunes $n_{\mathrm{g}}$ on the island. Unused gates are left uncolored. (c) False-color SEM of the nanowire before deposition of the top gates, showing the InAs core (orange) and the aluminum shell (blue). (d) Effective circuit diagram of the qubit. The weak-link with Josephson potential $U(\phi)$ is shunted by the island capacitance $C_{\mathrm{s}}$, $V_{\mathrm{g}}$ tunes $n_{\mathrm{g}}$, and $V_{\mathrm{j}}$ tunes the transparancy of the junction.}
\end{figure}
\footnotetext[4]{Uilhoorn, W. et. al. (2019). Parity lifetime of a proximitized semiconductor nanowire qubit in magnetic field. Manuscript in preparation.}

The measured gatemon is shown in Fig. \ref{fig:fig1}. The details of the device and experimental setup are provided in Ref. \cite{Note4} so we highlight only the relevant features here. The device consists of a superconducting island coupled to ground via an Al/InAs/Al weak-link \cite{Krogstrup2015, DeLange2015, Larsen2015}. The weak-link (shown in Fig.\ref{fig:fig1}c) is defined by etching away $\sim 100$~nm of the aluminum covering the InAs nanowire. A quantum dot is formed in the junction due to band-bending or disorder \cite{Chang2015, Deng2016}. The junction is shunted by the island capacitance $C_{\mathrm{s}}$, which predominantly sets the charging energy $E_{\mathrm{c}}\approx 750$~MHz. Electrostatic gates tune both the transparency of the junction and the dimensionless offset charge $n_{\mathrm{g}} = C_{\mathrm{g}} V_{\mathrm{g}}/2e$ on the island. The gatemon is capacitively coupled to a NbTiN $\lambda/2$ coplanar waveguide resonator \cite{Kroll2019} in order to excite and readout the system using standard dispersive readout techniques \cite{Bianchetti2009}. 

\begin{figure}[t!]
\includegraphics[width=8.6cm]{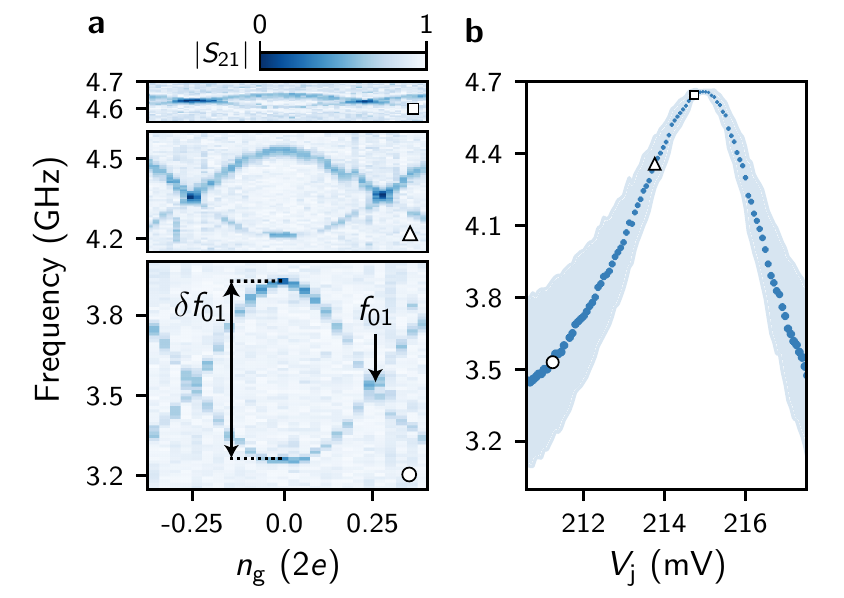}
\caption{\label{fig:fig2} Evolution of the qubit frequency and charge dispersion as a function of $V_{\mathrm{j}}$. (a) Normalized two-tone spectroscopy measurements of the $0\rightarrow 1$ transition versus the offset charge tuned by $V_{\mathrm{g}}$, measured at three successive values of $V_{\mathrm{j}}$: 211.2, 213.8, and 214.7 mV. (b) Extracted $f_{01}$ (markers) and $\delta f_{01}$ (shading and marker size) versus $V_{\mathrm{j}}$. Open markers indicate the positions of panel (a).}
\end{figure}

\footnotetext[5]{Measurements are integrated for 50~ms while quasiparticle poisoning takes place on a timescale of $\sim 100$~$\mu$s \cite{Note4}}
We measure the dependence of qubit's ground to first excited state transition frequency on the offset charge on the island [$f_{01}(n_{\mathrm{g}})$] using two-tone spectroscopy as shown in Fig. \ref{fig:fig2}a. Each measurement results in two sinusoidal curves shifted by half a period, belonging to qubit transitions for even and odd island parity. Their simultaneous detection is due to quasiparticle poisoning on timescales faster than the measurements \cite{Schreier2008, Note5}. We define the qubit frequency $f_{01}$ as the point of charge degeneracy between even and odd island parity and the charge dispersion $\delta f_{01}$ as the maximal frequency difference between the two parity states, reflecting the maximal energy cost of charging the island with an additional electron.

Figure~\ref{fig:fig2}a also demonstrates the behavior of $\{f_{01}, \delta f_{01}\}$ at different $V_{\mathrm{j}}$ near full depletion of the junction. In the lowest panel, we observe $f_{01}=3.539~\mathrm{GHz}$ and $\delta f_{01}=679~\mathrm{MHz}$ at $V_{\mathrm{j}}=211.2~\mathrm{mV}$. As $V_{\mathrm{j}}$ is increased, in the middle and top panel of Fig.~\ref{fig:fig2}a, $f_{01}$ increases to $4.629~\mathrm{GHz}$ while $\delta f_{01}$ decreases to $39~\mathrm{MHz}$. Figure~\ref{fig:fig2}b summarizes the dependence of $f_{01}$ and $\delta f_{01}$ as a function of $V_{\mathrm{j}}$. We observe that the qubit frequency exhibits a peak, increasing by a factor of 1.35 before decreasing again. The rise in $f_{01}$ is accompanied by a strong decrease in $\delta f_{01}$, suppressing by almost two orders of magnitude at the peak. This behaviour is consistent with the presence of a quantum dot in the junction, which has been linked to peaks in the critical current that coincide with transparencies close to unity \cite{Spanton2017, Hart2019a}. 

\begin{figure}[t!]
\includegraphics[width=8.6cm]{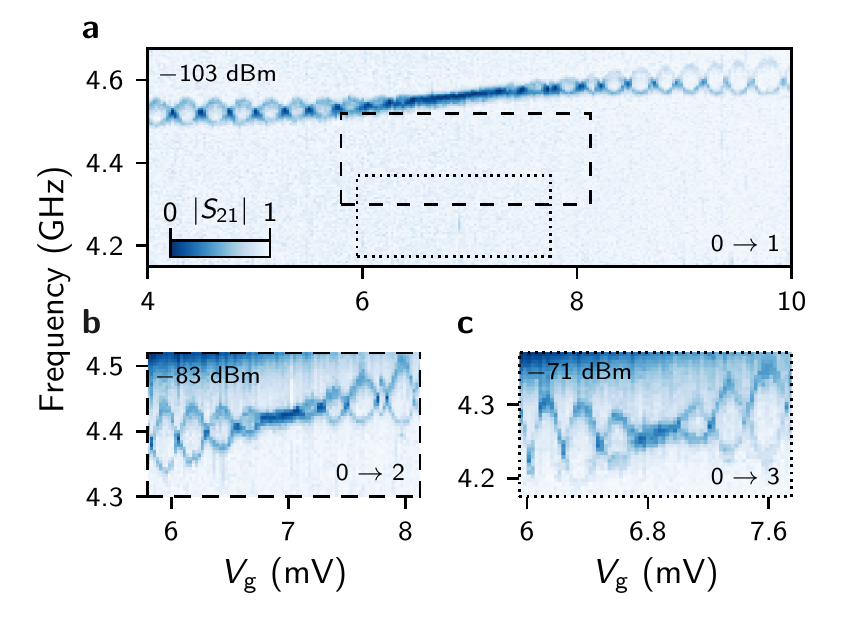}
\caption{\label{fig:fig3} Island gate dependence of the charge dispersion. (a) Normalized two-tone spectroscopy measurement of the $0\rightarrow1$ transition over a range of $V_{\mathrm{g}}$ encompassing many periods in offset charge. The charge dispersion suppresses down to the linewidth and subsequently recovers, while the qubit frequency increases over the entire gate range. (b)-(c) Multi-photon transitions $0\rightarrow2$ and $0\rightarrow3$, excited with increased driving powers. The transitions follow the same trends as panel (a), exhibiting an increased charge dispersion that still suppresses down to the linewidth at its minimum. Powers listed are at the sample input.}
\end{figure}

Due to finite stray capacitance, the transparency of the junction can also be tuned using the island gate. As shown in Fig. \ref{fig:fig3}a, we observe suppression of the charge dispersion by tuning the island gate voltage $V_{\mathrm{g}}$ with $V_{\mathrm{j}}$ fixed at a value where the charge dispersion is already close to the qubit linewidth $\gamma_{01} \approx 10 \ \mathrm{MHz}$ \footnote{These measurements are performed in a different cooldown of the same device.}. We note that $\delta f_{01}$ can no longer be discerned below $\gamma_{01}$ since the two parity transitions start to overlap. 

We can probe the suppression of Coulomb oscillations to below the limit set by $\gamma_{01}$ by measuring the charge dispersion of higher-order qubit transitions, which have a rapidly increasing charge dispersion $\delta f_{0n}$ \cite{Averin1999, Koch2007}. We repeat the measurement for increased driving powers in order to excite higher order qubit transitions. As shown in Fig. \ref{fig:fig3}b-c, the $0\rightarrow2$ and $0\rightarrow3$ multi-photon transitions indeed exhibit larger charge dispersion than the $0\rightarrow1$ transition. Even these larger charge dispersions vanish down to the linewidth $\gamma_{0n}$, indicating a particularly strong suppression. 

Beyond the remarkable suppression of the charge dispersion, we note that $\delta f_{01}$ does not depend monotonically on $f_{01}$. The charge dispersion of the $0\rightarrow1$ transition is suppressed down to the linewidth over several periods, while the qubit frequency slowly increases over the entire range of $V_{\mathrm{g}}$. Such a dependence cannot occur for superconducting tunnel junctions or a single mode superconducting quantum point contact (SQPC) \cite{Beenakker2005}, where larger larger qubit frequencies always result in lower charge dispersions \cite{Koch2007, Averin1999}. This behavior is the result of the quantum dot in the junction, in which case the charge dispersion need not be a monotonic function of the qubit frequency, as we explain below.

\footnotetext[6]{See the Supplemental Material for details of the data extraction, modeling of the ITLZ tunneling, the effective resonant level model, the fitting procedure and the estimation of the junction transparencies.} 

\begin{figure*}[hbt!]
\includegraphics[width=17.2cm]{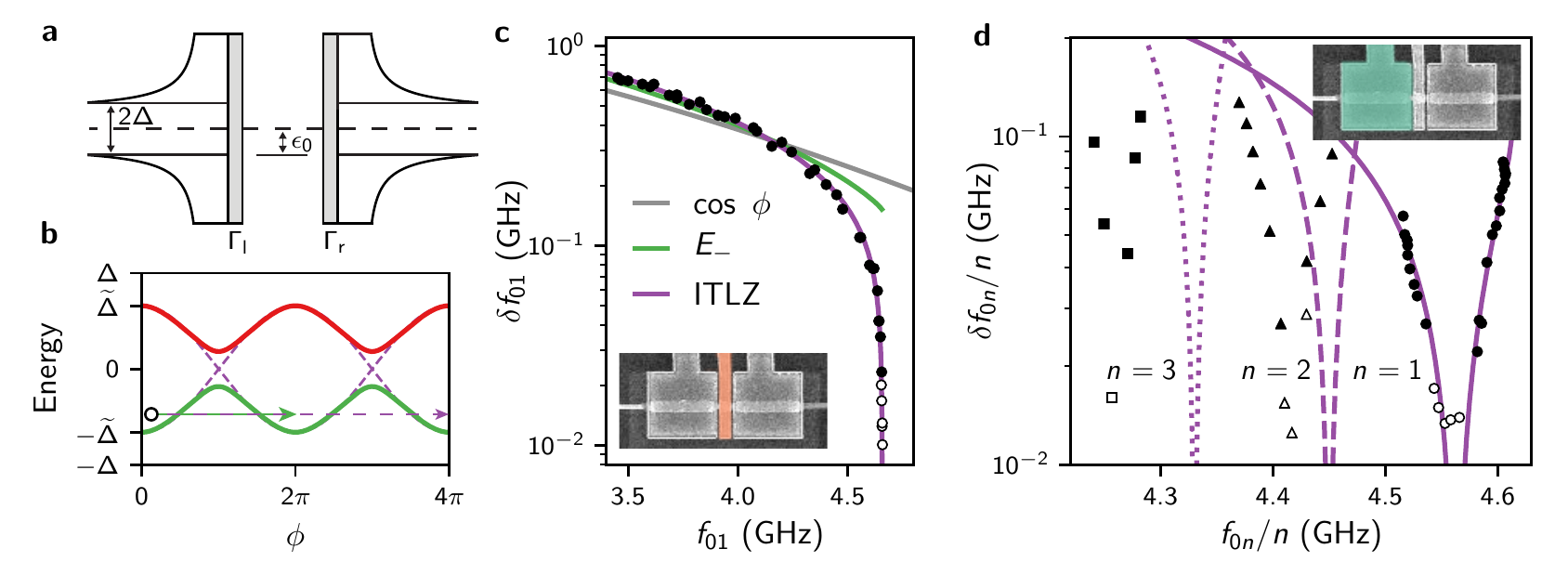}
\caption{\label{fig:fig4_new} Suppression of charge dispersion mediated by a resonant level. (a) Schematic depiction of a resonant level coupled to two identical superconducting leads. (b) Calculated energy-phase dependence of the ABS in the resonant level model for $\widetilde{D} = 0.9$ (solid) and $\widetilde{D} = 1$ (dashed) with $\Gamma = \Delta$. Arrows indicate the available quantum tunneling trajectories for the two cases. (c) Extracted qubit frequency versus charge dispersion measured in Fig. \ref{fig:fig2} by varying $V_{\mathrm{j}}$. Solid lines show fits using three models of $U(\phi)$: a sinusoidal potential, the negative energy ABS branch of the resonant level model, and a potential considering ITLZ tunneling between both ABS branches. (d) Extracted qubit frequency and charge dispersion of the first three transmon transitions measured as a function of $V_{\mathrm{g}}$ shown in Fig. \ref{fig:fig3}. The solid line shows a fit of the $0\rightarrow1$ transition with the ITLZ model, and the dashed and dotted lines show the resulting $0\rightarrow2$ and $0\rightarrow3$ transitions respectively. Open markers denote an upper bound on the charge dispersion based on the linewidth when $\delta f_{0n} \leq \gamma_{0n}$ and are not included in the fits.}
\end{figure*}

We develop a quantitative understanding of the device using a simplified model of a quantum dot between two superconducting leads known as the resonant level model \cite{ISI:A1989AU87800013, Beenakker2001, Devyatov1997, Golubov2004}. As shown in Fig. \ref{fig:fig4_new}a, we consider the presence of a single spin-degenerate level in the junction. The level has an energy $\epsilon_{\mathrm{0}}$ relative to the Fermi level, and is coupled to two identical superconductors with superconducting gap $\Delta$ via the (spin-degenerate) tunnel rates $\Gamma_{\mathrm{l}}$ and $\Gamma_{\mathrm{r}}$. Our simple model does not include the electron-electron interactions of the quantum dot. The potential of the junction $U(\phi)$ is determined by the energies of a single pair of spin degenerate ABS (shown in Fig. \ref{fig:fig4_new}b). Their energies have to be calculated numerically for general parameter values but can be expressed analytically in certain limits \cite{Devyatov1997, Golubov2004}:
\begin{equation}
\begin{aligned}
E_{\pm}(\phi) &= \pm \widetilde{\Delta} \sqrt{1-\widetilde{D} \sin^2 \phi/2}, \quad \widetilde{D} = \frac{4\Gamma_{\mathrm{l}}\Gamma_{\mathrm{r}}}{\epsilon_{\mathrm{0}}^2+\Gamma^2}, \\
\widetilde{\Delta} & =\begin{cases}
\Delta, & \text{if $\Gamma \gg \Delta, \epsilon_{\mathrm{0}}$}\\
  \Gamma& \text{if $\Gamma \ll \Delta$ and $\epsilon_{\mathrm{0}}=0$},
  \end{cases}
\end{aligned}
\label{eq:eqresonant}
\end{equation}
where $\Gamma = \Gamma_{\mathrm{l}}+\Gamma_{\mathrm{r}}$. 
Here the ABS take on the same functional form as for an SQPC \cite{Beenakker1991} but with an effective superconducting gap $\widetilde{\Delta} < \Delta$. The form of the effective junction transparency $\widetilde{D}$ also explicitly reflects a Breit-Wigner type resonant tunneling process, maximized for equal tunnel rates ($\delta\Gamma = \vert \Gamma_{\mathrm{l}} - \Gamma_{\mathrm{r}} \vert = 0$) and particle-hole symmetry ($\epsilon_{\mathrm{0}} = 0$). 

We now discuss the expected behavior of charge dispersion within this model. Under the typical assumptions of low to moderate values of $\widetilde{D}$ and $\widetilde{\Delta} \gg E_{\mathrm{c}}, k_B T$, only the ground state of the junction is occupied so that charge transfer occurs through $E_-$. In this regime, charge dispersion is exponentially suppressed in $\widetilde{\Delta} \widetilde{D}/E_{\mathrm{c}}$, comparable to the case of tunnel junctions and governed by tunneling of the phase under the potential barrier of $E_-$ \cite{Averin1985, Koch2007}.
As $\widetilde{D} \rightarrow 1$, however, the energy gap between the ABS vanishes. Due to ITLZ tunneling, the probability amplitude for the quantum tunneling trajectory to stay in the lower ABS branch vanishes linearly with the reflection amplitude $\sqrt{1-\widetilde{D}}$ \cite{Averin1999, Ivanov1998}. As a consequence $2\pi$-tunneling processes are suppressed, and so is the charge dispersion. When $\widetilde{D}=1$, the charge dispersion eventually saturates to a small value set by tunneling through a $4\pi$-wide potential barrier given by $\widetilde{\Delta}\cos{\phi/2}$.

Based on the discussion above, we fit the measured dependence of $\{f_{0n}, \delta f_{0n}\}$ using three junction models: a sinusoidal potential, a potential considering only the $E_-$ ABS branch of the resonant level model, and a potential including ITLZ tunneling. The numerical details of the procedure are described in the supplementary information \cite{Note6}. In Fig. \ref{fig:fig4_new}c, we plot the measured data for the dependence of $\delta f_{01}$ on $f_{01}$ while $V_{\mathrm{j}}$ is changed. In order to fit the data we assume that $V_{\mathrm{j}}$ tunes only $\epsilon_{\mathrm{0}}$ while $\Gamma_{\mathrm{l}} = \Gamma_{\mathrm{r}}$ are held constant. Furthermore, we fix $\Delta = 53 \ \mathrm{GHz}$ based on DC transport experiments on similar nanowires \cite{Deng2016}. The model based on a sinusoidal potential, which describes a tunnel junction with many low-transmission channels, is completely inconsistent with our data. Including only the presence of $E_-$ results in a fit that matches the initial decrease in $\delta f_{01}$ but is unable to capture the rapid suppression of the charge dispersion at the peak in qubit frequency. The model including ITLZ tunneling accurately describes the full range of data, requiring transparencies close to unity \cite{Note6}. We find that $\Gamma = 23 \ \mathrm{GHz}$, which gives an effective gap $\widetilde{\Delta} = 16 \ \mathrm{GHz}$ at the point of maximal suppression. The data and fit clearly demonstrate reaching the diabatic regime of ITLZ tunneling.

We additionally use the model including ITLZ tunneling to fit the $V_{\mathrm{g}}$ dependence of $\{f_{0n}, \delta f_{0n}\}$ in Fig. \ref{fig:fig4_new}d. Based on the position of the island gate to the left of the junction as well as screening by the junction gate, we assume that $V_{\mathrm{g}}$ tunes only $\Gamma_{\mathrm{l}}$ with all other parameters held constant. The resulting fit matches the characteristic shape of the data, showing strong suppression when $\delta\Gamma = 0$ and reproducing the non-monotonic relationship between qubit frequency and dispersion. The measurements also show that the anharmonicity $\alpha = f_{12} - f_{01}$ remains finite for all $\widetilde{D}$, essential for operation as a qubit. While the fit is excellent for the $0\rightarrow1$ transition, it requires a significantly lower superconducting gap $\Delta = 18.6~\mathrm{GHz}$. 
Additionally, the predicted qubit anharmonicities (indicated by the lines in Fig. \ref{fig:fig4_new}d) are lower than the measured anharmonicities, while the shapes of the curves remain accurate. This systematic deviation indicates that the underlying junction potential might be shallower than captured by our model. We speculate that the discrepancies in $\Delta$ and the anharmonicities could be due to omitting the electron-electron interactions of the quantum dot, which has previously been found to suppress the critical current and alter the energy-phase dependence of the ABS \cite{Vecino2003, Lim2008, Martin-Rodero2011, Hart2019a}. 

\begin{figure}[t!]
\includegraphics[width=8.6cm]{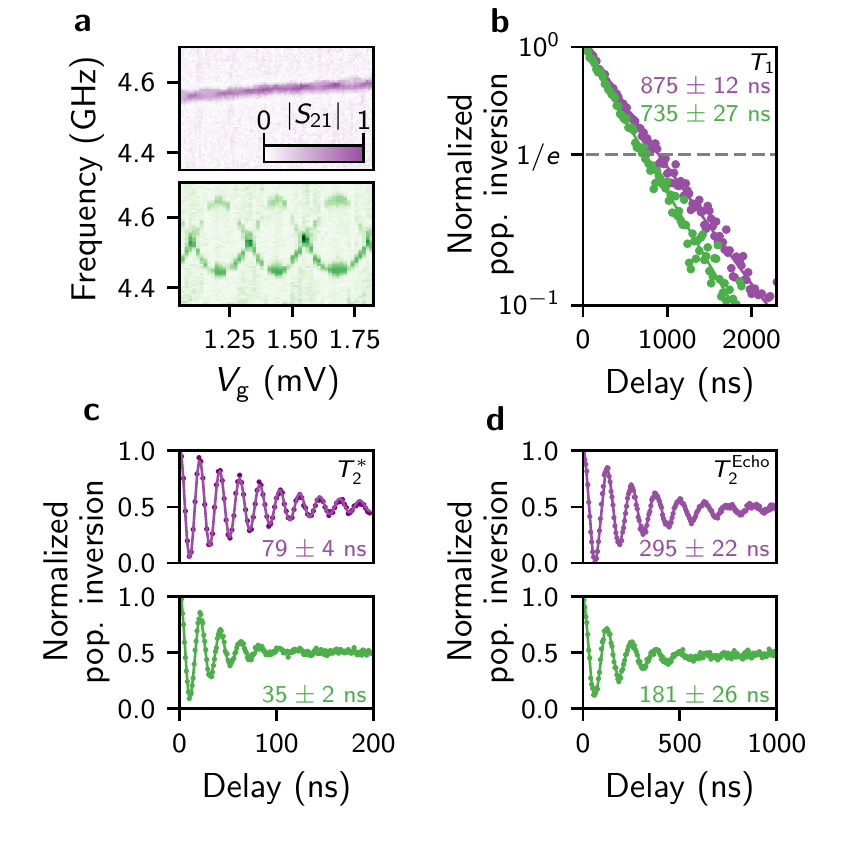}
\caption{\label{fig:fig4} Time-resolved qubit measurements for $\widetilde{D} \rightarrow 1$ (purple) and $\widetilde{D} \approx 0.8$ (green) in the junction measured at equal qubit frequency. (a) Two-tone spectroscopy measurements for the two cases. The remaining panels show measurements of $T_1$ (b), $T_2^{*}$ (c) and $T_2^{\mathrm{Echo}}$ (d).}
\end{figure}

Finally, we investigate the qubit's relaxation and coherence times in the presence and absence of resonant tunneling. Shown in Fig. \ref{fig:fig4}, we compare two cases: strong ITLZ tunneling with vanishing charge dispersion, and essentially adiabatic behaviour with $\delta f_{01} \ \approx 200 \ \mathrm{MHz}$. We leverage the non-monotonic $\{f_{01}, \delta f_{01}\}$ dependence encoded by $V_{\mathrm{g}}$ to make this comparison at nominally equal transition frequency in the same device. We find that the suppression leads to a moderately enhanced $T_1$. However, we do not expect charging effects to have a large effect on $T_1$ since the measurements are performed at $n_{\mathrm{g}} = 0.5$ where relaxation processes should be mostly charge-insensitive. We find, however, that both $T_2^{*}$ and $T_2^{\mathrm{echo}}$ improve considerably for the case of vanishing charge dispersion, reflecting the drastic reduction in sensitivity to charge noise. This is similar to the situation in conventional transmon qubits, where the exponential suppression of charge dispersion in $E_{\mathrm{J}}/E_{\mathrm{c}}$ is also accompanied by a strong increase in coherence times \cite{Koch2007}. However, in order to achieve the same level of $\delta f_{01}$ suppression in a conventional transmon for the $E_{\mathrm{c}}$ of this device one would require $E_{\mathrm{J}}/E_{\mathrm{c}} \geq 30$ , whereas we are operating at an effective $E_{\mathrm{J}}/E_{\mathrm{c}} \approx 5$. Even in the limit of full suppression, however, both relaxation and coherence times are short compared to results achieved in other gatemons \cite{Luthi2018}. We attribute these lower coherence times to ineffective radiation shielding and the quality of dielectrics used, which can be improved in future devices.

In summary, we measure the suppression of charge dispersion in an OCS gatemon with a highly transparent junction. We develop a model of tunneling through a resonant level in the junction that agrees with the dependence of the charge dispersion on both gates and indicates that, through tuning the parameters of our resonant level properly, we reach near-unity transparencies in our device.
Furthermore, the observed rate of suppression of the charge dispersion obeys the scaling law dictated by ITLZ tunneling between ABS. Finally we demonstrate that the suppression improves the qubit's coherence, reflecting the strong decrease in charge sensitivity. Independent research paralleling our own reports similar spectroscopic measurements on a full-shell gatemon with a DC transport lead \cite{Kringhoj2019}.

The vanishing of charging effects investigated here has implications for the design of hybrid circuits incorporating ballistic Josephson junctions \cite{Chauvin2007, Bretheau2013, Pekker2013, Hays2018, Tosi2019, Wang2019}. In particular, this vanishing may have positive implications for future gatemons \cite{DeLange2015,Larsen2015, Kringhoj2018, Serniak2019, Luthi2018}. The guaranteed vanishing of charge sensitivity for $\widetilde{D}\rightarrow 1$ while the anharmonicity remains finite places much less stringent requirements on $E_{\mathrm{c}}$ compared to other transmon implementations, allowing for faster qubit manipulation and strongly reducing the qubit's physical footprint. Finally, the natural magnetic field compatibility of S-QD-S transmons sets the stage for detecting and manipulating Majorana zero modes \cite{Hassler2011, Ginossar2014, Pikulin2019}. 

\begin{acknowledgments}
We acknowledge fruitful discussions with A. Antipov, A. Kringh\o j, J. Kroll, T.W. Larsen, R. Lutchyn, K. Petersson, W. Pfaff, D. Pikulin, and A. Proutski. We further thank Jasper van Veen for placing the nanowire, and Will Oliver for providing us with a traveling wave parametric amplifier. This work is part of the research project Scalable circuits of Majorana qubits with topological protection (i39, SCMQ) with project number 14SCMQ02, which is (partly) financed by the Dutch Research Council (NWO). It has further been supported by the Microsoft Quantum initiative.
\end{acknowledgments}

\bibliography{ms}

\end{document}

\preprint{APS/123-QED}

\title{Supplementary information for ``Observation of vanishing charge dispersion of a nearly-open superconducting island"}

\author{Arno Bargerbos}
\email[]{a.bargerbos@tudelft.nl}
\affiliation{QuTech and Kavli Institute of Nanoscience, Delft University of Technology, 2600 GA Delft, The Netherlands}
\author{Willemijn Uilhoorn}
\affiliation{QuTech and Kavli Institute of Nanoscience, Delft University of Technology, 2600 GA Delft, The Netherlands}
\author{Chung-Kai Yang}
\affiliation{Microsoft Quantum Lab Delft, 2600 GA Delft, The Netherlands}
\author{Peter Krogstrup}
\affiliation{Microsoft Quantum Materials Lab and Center for Quantum Devices, Niels Bohr Institute, University of Copenhagen, Kanalvej 7, 2800 Kongens Lyngby, Denmark}
\author{Leo P. Kouwenhoven}
\affiliation{QuTech and Kavli Institute of Nanoscience, Delft University of Technology, 2600 GA Delft, The Netherlands}
\affiliation{Microsoft Quantum Lab Delft, 2600 GA Delft, The Netherlands}
\author{Gijs de Lange}
\affiliation{Microsoft Quantum Lab Delft, 2600 GA Delft, The Netherlands}
\author{Bernard van Heck}
\affiliation{Microsoft Quantum Lab Delft, 2600 GA Delft, The Netherlands}
\author{Angela Kou}
\affiliation{Microsoft Quantum Lab Delft, 2600 GA Delft, The Netherlands}

\date{\today}% It is always \today, today,
       % but any date may be explicitly specified

\maketitle
\section{Data Extraction}
At each gate setting we measure the qubit transition frequency over at least one period in offset charge $n_g$. From this we extract the qubit frequency $f_{0n}$ and charge dispersion $\delta f_{0n}$ by applying a peak-finding algorithm to the raw two-tone spectroscopy data. The algorithm first smooths the data in frequency axis in order to combat noise, after which peaks are identified and fit with Lorentzian lineshapes in order to obtain their center frequency. For the gate voltage ranges in which two peaks are identified we take $\delta f_{0n}$ to be the local maximum in peak separation. Conversely, $f_{0n}$ is obtained from the regions where only a single peak is identified. In the regions of parameter space where $\delta f_{0n}$ is smaller than the qubit linewidth $\gamma_{0n}$ such that only a single peak can be discerned for any $n_g$ (open markers in Fig. 4 of the main text), we take the center frequency to be $f_{0n}$ and use the extracted linewidth of the Lorentzian lineshape as an upper bound for $\delta f_{0n}$.

\section{Modelling of the qubit}
\label{sec:supmodel}
In order to model the measured data we study the Hamiltonian of a capacitively shunted junction given by \begin{equation}
\hat{H} = 4 E_{\mathrm{c}} \left(\hat{n} - n_{\mathrm{g}}\right)^2 + U(\hat{\phi})
\label{eq:suphamiltonian}
\end{equation}
where $U(\hat{\phi})$ is the junction potential, $E_{\mathrm{c}}$ is the charging energy, $\hat{n}$ is the number of Cooper pairs that have traversed the junction, $n_{\mathrm{g}}$ is a dimensionless offset charge and $\hat{\phi}$ is the phase difference between the superconductors on either side of the junction. We obtain the qubit energy levels $E_n(n_g)$ and the corresponding qubit transitions $f_{ij}(n_g) = E_j(n_g) - E_i(n_g)$ through numerical diagonalization of the Hamiltonian, from which $f_{0n}$ and $\delta f_{0n}$ are calculated by evaluating the transitions at the appropriate offset charges.

We perform this procedure for three possible models for the junction potential: a sinusoidal potential as encountered in tunnel junctions, a potential considering only occupation of the $E_-$ ABS branch of the resonant level model, and a potential including ITLZ tunneling between the ABS of the resonant level model. In the case of the sinusoidal model we take $U(\hat{\phi}) = -E_{\mathrm{J}} \cos{\hat{\phi}}$, where we define an effective $E_{\mathrm{J}} \equiv \widetilde{\Delta} \widetilde{D} /4$ in order to compare the models on equal footing. For the model including only the $E_{-}(\hat{\phi})$ ABS branch we use $U(\hat{\phi}) = -\widetilde{\Delta} \sqrt{1-\widetilde{D} \sin^2 \hat{\phi}/2}$, while in order to include ITLZ tunneling between the $E_{\pm}(\hat{\phi})$ branches we follow the work of Ivanov and Feigel'man \cite{Feigelman1999a} and approximate the many-body superconducting system by an effective two-level system. This results in a junction potential given by 
\begin{equation}
U(\hat{\phi})    =
\widetilde{\Delta} \begin{pmatrix}
    \cos{\frac{\hat{\phi}}{2}} & \sqrt{1-\widetilde{D}}\sin{\frac{\hat{\phi}}{2}} \\[4pt]
    \sqrt{1-\widetilde{D}}\sin{\frac{\hat{\phi}}{2}}  & -\cos{\frac{\hat{\phi}}{2}} 
\end{pmatrix}
\label{eq:feigelman}
\end{equation}
In the above $\widetilde{\Delta}$ and $\widetilde{D}$ are effective parameters resulting from the underlying quantum dot parameters discussed in section \ref{sec:QD}.

\begin{figure}[t!]
\includegraphics[width=12.9cm]{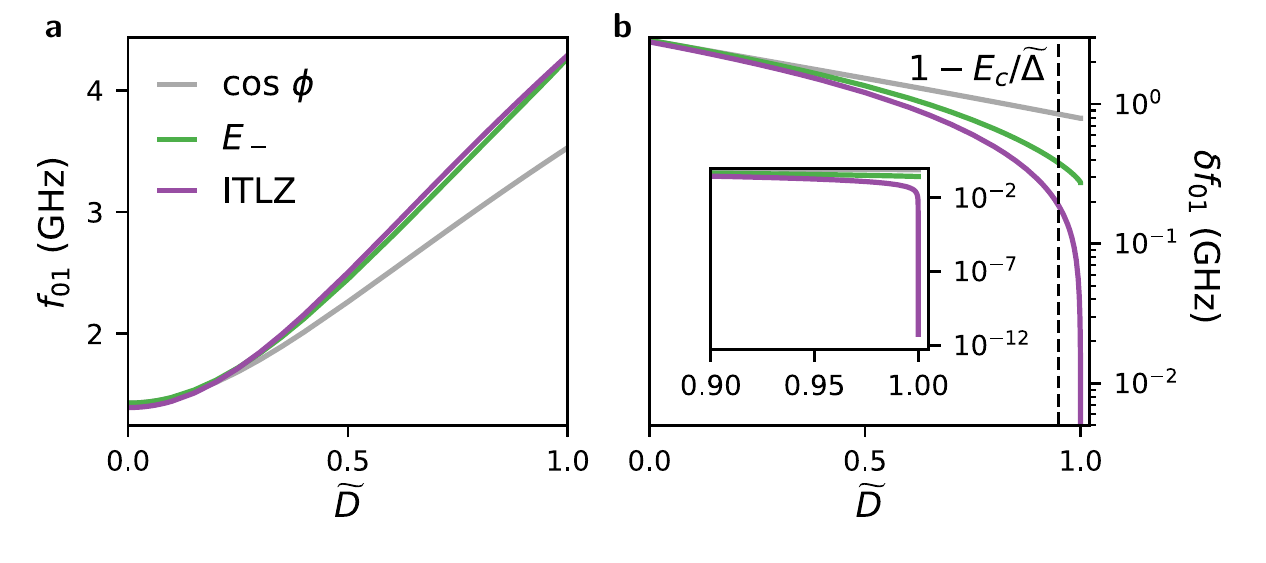}% Here is how to import EPS art
\caption{\label{fig:supaverin}  Qubit frequency (panel a) and charge dispersion (panel b) evaluated for three different models: a sinusoidal potential, the negative energy ABS branch of the resonant level model, and a potential considering ITLZ tunneling between both ABS branches. All models are evaluated for fixed parameter values $\widetilde{\Delta} = 14$~GHz and $E_c = 715$~MHz. The dashed line indicates the crossover value $1-E_c/\widetilde{\Delta}$. Inset of panel b: Zoom in of the behaviour near $D=1$, where the ITLZ model exhibits rapid suppression in charge dispersion down to a small value set by tunneling through a potential barrier $\widetilde{\Delta} \cos \phi/2$.}
\end{figure}

The results of this procedure are demonstrated in Fig. \ref{fig:supaverin}, which shows how $f_{01}$ and $\delta f_{01}$ depend on $\widetilde{D}$ for the three models. The sinusoidal model reproduces the expected results of the conventional tunnel junction transmon, exhibiting exponential suppression of $\delta f_{01}$ for large values of $\widetilde{D}$ and thus $E_{\mathrm{J}}/E_{\mathrm{c}}$. The $E_{-}(\phi)$ model exhibits similar behaviour up to moderate values of $\widetilde{D}$, after which an enhanced suppression of $\delta f_{01}$ takes place due to the increased height of the potential compared to the sinusoidal model. Finally, the model including ITLZ tunneling shows comparable behaviour to $E_{-}(\phi)$ up to large values of $\widetilde{D}$, after which a much more rapid decrease in charge dispersion takes place. This reproduces the scaling law predicted for ballistic Josephson junctions \cite{Averin1999}. We note that the cross-over value of $\widetilde{D}$ between the adiabatic regime well-described by only $E_-$ and the diabatic regime including ITLZ tunneling is approximately given by $\widetilde{D} = 1-E_c/\widetilde{\Delta}$, where the rate of phase evolution becomes comparable to the energy gap between the ABS \cite{Feigelman1999}. 

\section{Resonant Level Model}
\label{sec:QD}
\begin{figure}[t!]
\includegraphics[width=12.9cm]{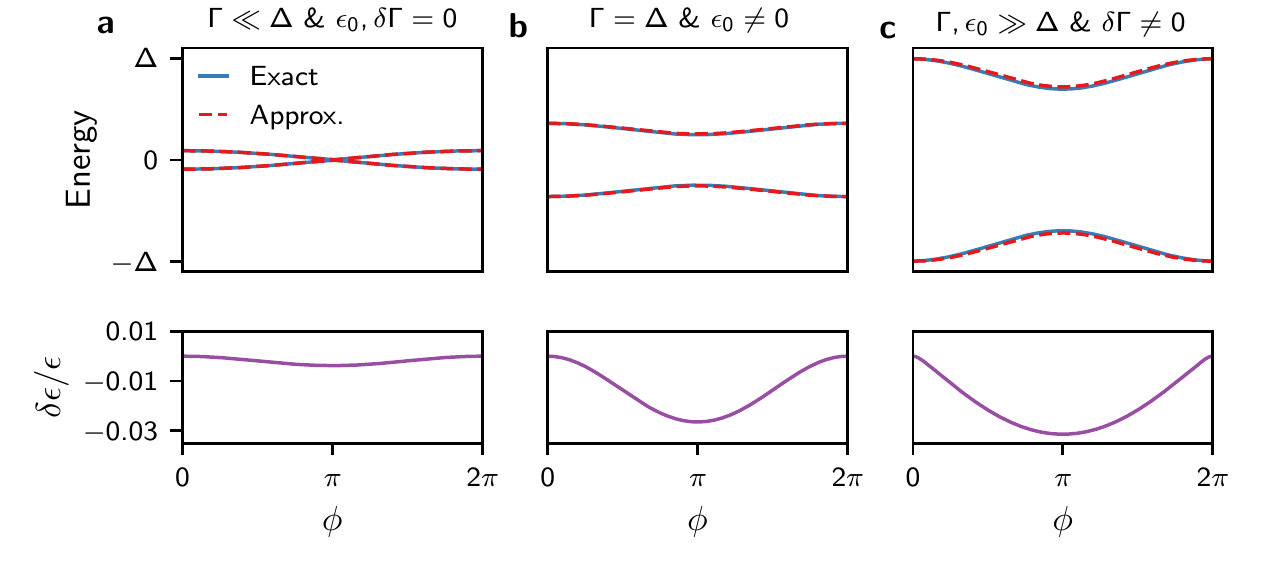}% Here is how to import EPS art
\caption{\label{fig:QDerr} Comparison between the exact and approximate solutions for the ABS of the resonant level model. Top half shows the explicit energy levels for both solutions while the bottom half shows the relative error of the approximate model. Panel a shows the comparison in the regime of weak coupling ($\Gamma \ll \Delta)$ on resonance ($\epsilon_0 = 0$) with symmetric barriers ($\delta\Gamma = 0$). Panel b shows the regime of moderate coupling in which $\Gamma_{\mathrm{l}}=\Gamma_{\mathrm{r}} = \Delta$ and where the level is off-resonant ($\epsilon_0 \neq 0$). Panel c shows the regime of far off-resonant strong coupling with asymmetric tunneling rates, in which $\Gamma, \epsilon_0 \gg \Delta$ and $\delta\Gamma \neq 0$.}
\end{figure}

In order to develop a quantitative understanding of our device we study a simplified model of a quantum dot between two superconducting leads known as the resonant level model \cite{Beenakker2001}. Depicted in Fig. 4a of the main text, it considers the presence of a single spin-degenerate level in the junction with an energy $\epsilon_{\mathrm{0}}$ relative to the Fermi level, coupled to two identical superconductors with superconducting gap $\Delta$ via the spin-degenerate tunnel rates $\Gamma_{\mathrm{l}}$ and $\Gamma_{\mathrm{r}}$. Its discrete energy spectrum follows from the solutions $\epsilon \in (0,\Delta)$ of the equation
\begin{equation}
\left(\Delta^2-\epsilon^2)(\epsilon^2-\epsilon_0^2-\Gamma^2\right) + 4\Delta^2\Gamma_{\mathrm{l}}\Gamma_{\mathrm{r}} \sin{}^2(\phi/2) + 2\Gamma \epsilon^2(\Delta^2-\epsilon^2)^{1/2} = 0
\label{eq:supbeenakker}
\end{equation}
where $\Gamma = \Gamma_{\mathrm{l}}+\Gamma_{\mathrm{r}}$. 
This equation can be solved numerically for general parameter values, resulting in a single pair of spin degenerate ABS $E_{\pm}(\phi)$. Furthermore, in certain limits its solution can be recovered analytically (given in Eq. 1 of the main text), which coincides with the eigenvalues of Eq. \ref{eq:feigelman}. However, we found these limits too constraining for the model to accurately describe our data. We therefore construct an approximate solution to Eq. \ref{eq:supbeenakker} based on the ABS energies $E_{\pm}(\phi)$ and transparency $\widetilde{D}$ given by Eq. 1 of the main text, whereas for $\widetilde{\Delta}$ we do not use the limiting values but instead solve Eq. \ref{eq:supbeenakker} for the bound state energy at $\phi = 0$. Shown in Fig. \ref{fig:QDerr}, we tested the validity of this approximation for a wide range of parameters by explicit comparison to the solutions of Eq. \ref{eq:supbeenakker}. The effective spectrum closely resembles the exact solutions over a wide range of parameters, with relative errors on the order of a few percent even in the regime $\Gamma \approx \Delta$. We therefore argue that the effective model of Eq. \ref{eq:feigelman} should accurately describe the phenomenology of the resonant level junction. A more detailed description of the model and its derivation can be found in \cite{Kringhoj2019a}.

\begin{figure}[t!]
\includegraphics[width=12.9cm]{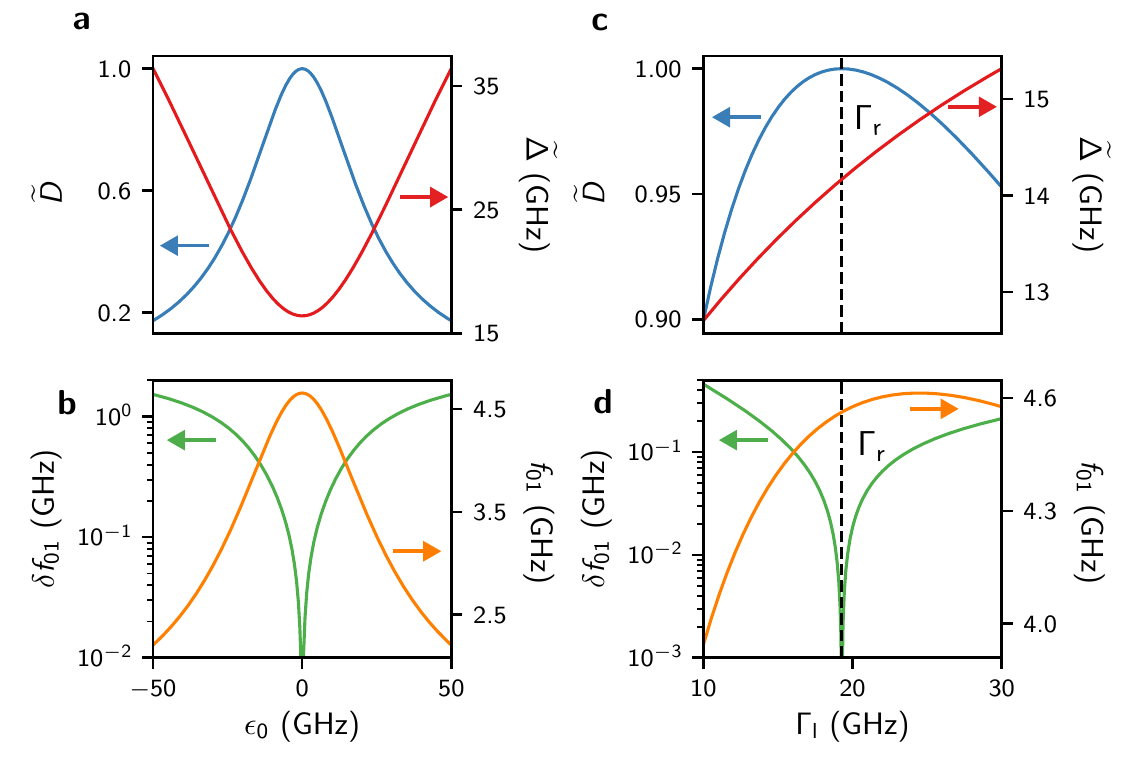}% Here is how to import EPS art
\caption{\label{fig:fig4d} Effective junction and qubit parameters as a function of resonant level parameters. Panel a (b) shows how $\widetilde{D}$ and $\widetilde{\Delta}$ ($f_{01}$ and $\delta f_{01}$) behave as a function of $\epsilon_0$ for $\delta\Gamma = 0$, panel c (d) shows how these parameters behave as a function of $\Gamma_{\mathrm{l}}$ for $\epsilon_0 =0$ and $\delta \Gamma \neq 0$.}
\end{figure}

Having constructed the effective parameters $\widetilde{D}$ and $\widetilde{\Delta}$, we now show how the ITLZ model introduced in section \ref{sec:supmodel} behaves as a function of the underlying quantum dot parameters. We do this for the parameter values that resulted in the best fits to the measured data, shown in Fig. 4 of the main text. In Fig. \ref{fig:fig4d} we study the effect of varying $\epsilon_0$ at $\Gamma_{\mathrm{l}} = \Gamma_{\mathrm{r}}$. It results in a weak dependence of $\widetilde{\Delta}$, which is minimal when $\epsilon_0 = 0$. This coincides coincides with a maximum in $\widetilde{D}$, as given by Eq. 1 of the main text. Shown in panel b this translates into a qubit frequency that is maximal at $\epsilon_0 = 0$, coinciding with a minimum in charge dispersion. Panels c and d in turn show the dependence on $\Gamma_{\mathrm{l}}$ at fixed $\Gamma_{\mathrm{r}}$ with $\epsilon_0 = 0$. We find that $\widetilde{\Delta}$ is a monotonically increasing function of $\Gamma_{\mathrm{l}}$, whereas $\widetilde{D}$ is maximal when the asymmetry $\delta\Gamma = \vert \Gamma_{\mathrm{l}}-\Gamma_{\mathrm{r}} \vert$ is minimized. Panel d shows that the relative rapidity at which $\widetilde{\Delta}$ and $\widetilde{D}$ change around $\delta\Gamma = 0$ can result in a situation where the maximal qubit frequency does not coincide with minimal charge dispersion. We believe this effect to be the origin of the non-monotonic dependence between the qubit frequency and the charge dispersion seen in figures 3 and 4d of the main text.

\section{Fitting Routine}
We fit the measured relationships between $\{f_{0n}, \delta f_{0n}\}$ using the models developed in section \ref{sec:supmodel}. For the data measured as a function of $V_{\mathrm{j}}$, shown in Fig. 4c of the main text, we assume that only $\epsilon_{\mathrm{0}}$ is varied. The remaining remaining parameters $\Delta$, $E_{\mathrm{c}}$, $\Gamma_{\mathrm{l}}$, and $\Gamma_{\mathrm{r}}$ are taken to be independent. Additionally, we fix $\Delta = 53 \ \mathrm{GHz}$ based on DC transport experiments on similar nanowires \cite{Deng2016} and we assume that $\Gamma_{\mathrm{l}} = \Gamma_{\mathrm{r}}$. We then apply a fitting routine in which for each set of parameter values a range of $\{\widetilde{\Delta}, \widetilde{{D}}\}$ is generated from the effective resonant level ABS potential for a large range of $\epsilon_{\mathrm{0}}$. These effective parameters are then used in the three different junction potentials of section \ref{sec:supmodel}, resulting in a set of calculated values for $\{f_{01}, \delta f_{01}\}$ that can be compared to the measured values. The best fit to the data for each model is obtained through the standard method of least-squares.

For the extracted relationships between $\{f_{0n}, \delta f_{0n}\}$ as a function of $V_{\mathrm{g}}$, shown in Fig. 4d of the main text, this procedure is slightly modified. We now assume that only $\Gamma_{\mathrm{l}}$ is a function of $V_{\mathrm{g}}$, with $\Delta$, $E_{\mathrm{c}}$, $\epsilon_{\mathrm{0}}$, and $\Gamma_{\mathrm{r}}$ taken to be independent. For simplicity we fix $\epsilon_{\mathrm{0}} = 0$. We note that, in order to obtain a good fit for the data versus $V_{\mathrm{g}}$, $\Delta$ needed to enter as a free parameter. In addition, a good fit could not be found simultaneously for all three measured transitions with a single set of parameters. Instead we only fit the $\{f_{01}, \delta f_{01}\}$ data to the model, resulting in a qualitatively satisfying fit for the $01$ transition. However, the fit suggests a value of superconducting gap $\Delta = 18.6$~GHz, much smaller than the value measured in DC experiments \cite{Deng2016}. Moreover, the fit does not manage to capture the position of the higher order transitions. As discussed in the main text, we attribute this parameter discrepancy as well as the inability to fit all three transitions to possible modifications in the shape of the potential originating from the lack of electron-electron interactions in the model.

\section{Estimating transparencies}
\begin{figure}[t!]
\includegraphics[width=12.9cm]{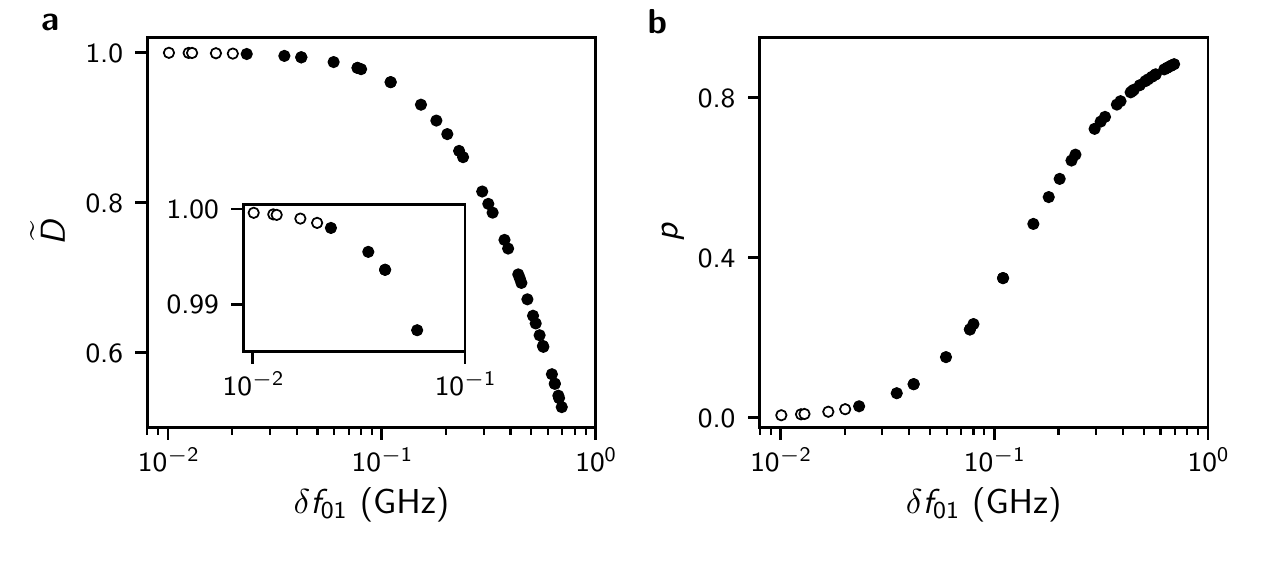}% Here is how to import EPS art
\caption{\label{fig:ts} (a) Extracted $\widetilde{D}$ for the values of $\delta f_{01}$ measured as a function of $V_{\mathrm{j}}$, with the inset showing the behaviour near $\widetilde{D}=1$. (b) Asymptotic probability for the ABS to remain in the $E_-$ branch as calculated from the extracted $\widetilde{D}$ and $\widetilde{\Delta}$.}
\end{figure}

An estimate for the transparencies $\widetilde{D}$ realized the in the experiment can be obtained from the fits to the data. As illustrated in Fig. \ref{fig:fig4d}, each numerically calculated value of $\delta f_{0n}$ corresponds to a value of $\widetilde{D}$, and one can therefore infer the values of $\widetilde{D}$ by matching the measured values of $\delta f_{0n}$ to the numerical values. We emphasize that these values are model and parameter dependent, and are therefore only an estimate. Shown in Fig. \ref{fig:ts}a, we find that by varying $V_{\mathrm{j}}$ transparencies between 0.5 and 1 are attained, with the largest transparency based on a distinguishable charge dispersion (filled markers) being 0.998 and the largest value based on the qubit linewidth $\gamma_{01}$ (open markers) being 0.9996. Finally in panel b we show the asymptotic probability $p$ of the ABS remaining in the ground state as calculated from the extracted $\widetilde{D}$ and $\widetilde{\Delta}$ \cite{Averin1999}. This illustrates that the suppression of charge dispersion coincides with the vanishing of $p$. Furthermore, it shows that sizeable ITLZ probabilities are obtained over a wide range of the measured values, robust to small changes in fit parameters. We do not repeat this procedure for the data obtained by varying $V_{\mathrm{g}}$, given the unsatisfactory fit to the data.

\bibliography{supplement}%